\documentstyle{article} 
\author{Horacio Alberto FALOMIR$^1$ \and Ricardo Enrique GAMBOA SARAVI
$^1$ \and Mar\'{\i}a Amelia MUSCHIETTI$^2$ \and Eve Mariel SANTANGELO$^1$
\and ~~~~~~~~~~~~~~Jorge Eduardo SOLOMIN$^2$~~~~~~~~~~~~~~ \\ \hfill\\
\and {\Large Facultad de Ciencias Exactas} \\ {\Large Universidad Nacional
de La Plata}\\ $^1$Departamento de F\'{\i}sica\\ $^2$Departamento de
Matem\'atica}

\title{On the relation between determinants and Green functions of
elliptic operators with local boundary conditions.  \thanks{Partially
supported by CONICET and Fundaci\'on Antorchas, Argentina. } } \date{ }
 \def\dfrac#1#2{{\displaystyle
{#1 \over #2}}}  \def\QATOP#1#2{{#1 \atop
#2}} 
  
\catcode`\@=11
\def\qed{\ifhmode\unskip\nobreak\fi\ifmmode\ifinner\else\hskip5\p@\fi\fi
 \hbox{\hskip5\p@\vrule width4\p@ height6\p@ depth1.5\p@\hskip\p@}}

\begin{document}

\maketitle \begin{abstract} A formula relating quotients of determinants
of elliptic differential operators sharing their principal symbol, with
local boundary conditions, to the corresponding Green function is given. 

\bigskip
\centerline{\bf R\'esum\'e}

On \'etablit une formule reliant quotients des d\'eterminants des
op\'erateurs diff\'erentiels elliptiques qui partagent leur symbol
pricipal, avec des conditions au bord locales, et les fonctions de Green
associ\'ees.

\end{abstract} 
                   
 \eject 
 
\section{Introduction} 
 
Functional determinants have wide applications in Quantum and Statistical
Physics. A powerful tool to regularize such determinants in a gauge
invariant way, the so called $\zeta$-function method \cite{Haw}, is based
on Seeley's construction of complex powers of elliptic differential
operators. 
 
This construction has been largely studied and applied in the case of
boundaryless manifolds,(see, for instance, \cite{Annals} and references
therein). 
 
For manifolds with boundary, the study of complex powers was performed in
\cite{Seeley1,Seeley2} for the case of local boundary conditions, while
for the case of nonlocal conditions, this task is still in progress (see,
for example, \cite{SG}.)

The aim of this paper is to establish a relationship between determinants
of differential operators, under local elliptic boundary conditions, and
the corresponding Green functions, which involves a finite number of
Seeley's coefficients. (We have perfo rmed an application of such a
relationship to a concrete physical situation, involving Dirac operators,
in a previous paper \cite{porfin}, where no mathematical proofs were
included.)

For the sake of simplicity, only first order operators will be considered
in the following, although straightforward modifications would allow to
generalize this result to elliptic boundary problems of any order. 

\section{Seeley's complex powers and regularized determinants for local
elliptic boundary problems}

Let $D$ be a first order elliptic operator,

\begin{equation} \label{OP}D:{\cal C}^\infty (M,E)\rightarrow {\cal
C}^\infty (M,F), \end{equation} where $M$ is a bounded closed domain in
${\bf R}^\nu $
 with smooth boundary $\partial M$, and E and $F$ are $k$  -dimensional
complex vector bundles over $M,$ with a local boundary condition $B:
L^2(E/ \partial M) \rightarrow L^2(G)$ being $G$ a $r$-dimensional complex
vector bundle over $\partial M$, $r<k$. 
 
In a collar neighborhood of $\partial M$ in $M,$ we will take coordinates
$  \bar x=(x,t)$, with $t$ the inward normal coordinate and $x$ local
coordinates for $\partial M$, and conjugated variables $\bar \xi =(\xi
,\tau )$ in $T^{*}M$.

The elliptic boundary problem

\begin{equation} \label{BoundaryProblem}\left\{ \begin{array}{c} D\varphi
=0 \ \rm{ in }\ M \\ \\ B\varphi =f\ \rm{ on }\ \partial M \end{array}
\right.  \end{equation} is said to admit a cone of Agmon's directions if
there is a cone $\Lambda$ in the $\lambda$ complex plane such that

1) $\forall \bar x\in M$, $\forall \bar \xi \neq 0,$ $\Lambda $ contains
no eigenvalues of the matrix $\sigma _1(D)(\bar x,\bar \xi )$. 
 
and

2) $\forall \lambda \in \Lambda ,$ $\forall x\in \partial M,$ $\forall
g\in {\bf C}^{r}$, the initial value problem  $$ \begin{array}{c} \sigma
_1(D)(x,0;\xi,-i \partial _{t})\ u(t)=\lambda \ u(t) \\ \\ b(x)\ u(0) =g
\end{array} $$ has, for each $\xi \neq 0,$ a unique solution satisfying
$\lim \limits_{t\rightarrow \infty }\ u(t)=0$, being $ \sigma _1(D)$ the
principal symbol of $D$ and $b(x)$ such that $B (\phi) (x) = b(x)
\phi(x)$.

Henceforth, we assume the existence of an Agmon's cone $\Lambda $.
Moreover, we will consider only boundary conditions $B$ giving rise to a
discrete spectrum $sp(D_B)$, where $D_B$ denotes the closure of $D$ acting
on the sections $\phi \in {\cal C}^\in fty (M,E)$ satisfying $B \phi = 0$
on $\partial M$. Note that, this is always the case for elliptic boundary
problems unless $sp(D_B)$ is the whole complex plane. Now, for $\vert
\lambda \vert $ large enough, $sp(D_B)\cap \Lambda $ is empty, since there
is no $\lambda $ in $sp(\sigma _1(D_B))\cap \Lambda $.  Then, $sp(D_B)\cap
\Lambda $ is a finite set. 

For $\lambda \in \Lambda $ not in $sp(D_B)$, and

\begin{equation} \sigma (D- \lambda I ) = a_{0}(x,t;\xi,\tau;\lambda) +
a_{1}(x,t;\xi,\tau;\lambda), \label{a} \end{equation} with $a_l$
homogeneous of degree $l$ in $(\bar{\xi} , \lambda)$, an asymptotic
expansion of the symbol of $R(\lambda )=(D_B-\lambda I)^{-1}$ can be
explicitly given \cite {Seeley1}: 
 
\begin{equation} \label{AE}\sigma (R(\lambda ))\sim \sum_{j=o}^\infty
c_{-1-j}-\sum_{j=o}^\infty d_{-1-j} \end{equation} where the {\it Seeley
coefficients } $c_{-1-j}$ and $d_{-1-j}$ satisfy
 
\begin{equation} \label{7}\sum_{j=o}^\infty a_{1-j}\ \ \circ \
\sum_{j=0}^\infty c_{-1-j}=I, \end{equation} $\circ $ denoting the usual
composition of homogeneous symbols (see for instance
\cite{LibroAmarillo},) and \begin{equation} \label{9}\left\{
\begin{array}{c} \sigma ^{\prime }(D-\lambda )\ \circ \
\sum\limits_{j=o}^\infty d_{-1-j}=0 \\ \\ \sigma (B)\ \circ \
\sum\limits_{j=o}^\infty d_{-1-j}=\sigma (B)\ \circ \ \
\sum\limits_{j=0}^\infty c_{-1-j}\quad \rm{at}\ \ t=0 \\ \\ \lim
\limits_{t\rightarrow \infty }\ d_{-1-j}=0.  \end{array} \right. 
\end{equation} Here $\sigma ^{\prime }(D-\lambda I)$, the ``partial
symbol" of $D$ at the boundary, is defined as follows:  \begin{equation}
\sigma ^{\prime }(D-\lambda I)= \sum\limits_{j}^{}{a^{(j)}},
\end{equation} where

\begin{equation} a^{(j)} = a^{(j)}(x,t,\xi,-i\partial_{t},\lambda)
=\sum_{l-k=j}^{}{{t^k\over{k !}}
a_{l}^{(k)}(x,0,\xi,-i\partial_{t},\lambda)}, \end{equation} with
$a_{l}^{(k)}= \partial_{t}^{k} a_l $ and $a_l $ as in (\ref{a}). 

Note that condition 2) implies the existence and unicity of the solution
of (\ref{9}). 

Written in more detail, the first line in (\ref{9}) becomes \cite{Seeley1}
\begin{equation} \label{11}a^{(1)}d_{-1-j}+\sum_{\QATOP{l<j}{k-\vert
\alpha \vert -1-l=-j}} \frac{i^\alpha }{\alpha !}\frac{\partial ^\alpha
}{\partial \xi ^\alpha } a^{(k)}\frac{\partial ^\alpha }{\partial x^\alpha
}d_{-1-l}=0, \end{equation} while the second one is \begin{equation}
\label{12} b_0d_{-1-j}=b_{0}c_{-1-j}\vert _{t=0} .  \end{equation}
 
It is worth noticing that, although \begin{equation} \label{Aet}\sigma
(R(\lambda ))=\ \sum_{j=0}^\infty c_{-1-j}\ , \end{equation} is an
asymptotic expansion of $\sigma (R(\lambda )),$ the fundamental solution
of ($D_B-\lambda )$ obtained by Fourier transforming Eq.(\ref{Aet}) does
not in general satisfy the required boundary conditions. In fact, the
coefficients $d_{-1-j}$ are added to the expansion in order to correct
this deficiency. 
 
The coefficients $c_{-1-j}$ $(x,t;\xi ,\tau ;\lambda )$ and $ 
d_{-1-j}(x,t;\xi ,\tau ;\lambda )$ are meromorphic functions of $\lambda $
with poles at those points where $\det [\sigma _1(D-\lambda )(x,t;\xi
,\tau )]$ vanishes. The $c_{-1-j}$'s are homogeneous of degree $-1-j$ in (
$\xi ,\tau ,\lambda )$; the $d_{-1-j}$ 's are also homogeneous of degree
$-1-j,$ but in $( \frac 1t,\xi ,\tau ,\lambda)$ \cite{Seeley1}. 

From these coefficients we get an approximation to $(D_B-\lambda )^{-1}$,
a parametrix constructed as in \cite{Seeley1}
 
\begin{equation} \label{100}P_K(\lambda )=\sum_\varphi \psi \left[
\sum_{j=0}^KOp(\theta _2\ c_{-1-j})-\sum_{j=0}^KOp^{\prime }(\theta _{1\
}\tilde{d}_{-1-j})\right] \ \varphi , \end{equation} where $\varphi $ is a
partition of the unity, $\psi \equiv 1$ in $  Supp(\varphi )$, $\theta_1$
and $\theta_2$ cut-off functions for $\vert \xi \vert ^2+\vert \lambda
\vert ^2 \ge 1$ and $\chi (\vert \xi \vert ^2+\vert \tau \vert ^2+\vert
\lambda \vert ^2 \ge 1$ respectively, and

\begin{equation} \begin{array}{c} \displaystyle Op(\sigma )h(x,t)=\int
\sigma (x,t;\xi ,\tau )\ \hat h(\xi ,\tau )\ e^{i(x\xi +t\tau )}
\dfrac{d\xi }{(2\pi )^{\nu -1}}\ \dfrac{d\tau }{2\pi }, \\ \\ \\
\displaystyle Op^{\prime }(\sigma )h(x,t)=\int \int \tilde \sigma (x,t;\xi
,s)\ \tilde h(\xi ,s)\ e^{ix\xi }\dfrac{d\xi }{(2\pi )^{\nu -1}}\
\dfrac{ds}{2\pi }, \end{array} \end{equation} \\ where $\ \hat h(\xi ,\tau
)$ is defined in (\ref{Fourier}) and \begin{equation} \tilde h(\xi
,s)=\int \ h(x,s)\ e^{-ix\xi }\ dx.\ \end{equation} \\

Moreover, it can be proved  that, for $\lambda \in \Lambda ,$
 \begin{equation} \label{SG}\parallel R(\lambda )\parallel _{L^2}\leq
C\vert \lambda \vert ^{-1} \end{equation} with C a constant
\cite{Seeley1,Gil}. 
 
The estimate (\ref{SG}) allows for expressing the complex powers of $D_B$
as \begin{equation} \label{CP}D_B^z=\frac i{2\pi }\int_\Gamma \lambda ^z\
R(\lambda )\ d\lambda \end{equation} for $Re\ z<0$ , where $\Gamma $ is a
closed path lying in $\Lambda $, enclosing the spectrum of $D_B$
\cite{Seeley2}$.$ Note that such a curve $  \Gamma $ always exists for
$sp(D_B)\cap \Lambda $ finite. 
 
For $Re\ z\geq 0$ , one defines  \begin{equation} D_B^z=D^l\circ
D_B^{z-l}\ , \end{equation} for $l$ a positive integer such that $Re\
(z-l)<0$. 
 
If $Re(z)<-\nu $, the power $D_B^z$ is an integral operator with
continuous kernel $  J_z(x,t;y,s)$ and, consequently, it is trace class
(for an operator of order $\omega $, this is true if $Re(z)<-\frac \nu
\omega $). As a function of $z$, $Tr(D_B^z)$ can be extended to a
meromorphic function in the whole complex plane {\bf C}, with only simple
poles at $z=j-\nu ,\ j=0,1,2,...$ and vanishing residues when
$z=0,1,2,...$ (for an operator of order $\omega $ , there are only single
poles at $z=\frac{j-\nu }\omega ,\ j=0,1,2,...$, with vanishing residues
at $z=0,1,2,...$) \cite{Seeley2}. 

 The function $Tr(D_B^z)$ is usually called $\zeta _{(D_B)}(-z)$ because
of its similarity with the classical Riemann $\zeta $-function: if
$\{\lambda _j\}$ are the eigenvalues of $D_B$, $\{\lambda _j^z\}$ are the
eigenvalues of $D_B^z$; so $  Tr(D_B^z)=\sum \lambda _j^z$ when $D_B^z$ is
a trace class operator. 
 
A regularized determinant of $D_B$ can then be defined as \begin{equation}
\label{DR}Det\ (D_B)=\exp [-\frac d{dz}\ Tr\ (D_B^z)]\vert _{z=0} . 
\end{equation}
 
Now, let $D(\alpha )$ be a family of elliptic differential operators on
$M$ sharing their principal symbol and analytically depending on $\alpha
$. Let $B$ give rise to an elliptic boundary condition for all of them, in
such a way that $D(\alpha )_B$ is invertible and the boundary problems
they define have a common Agmon's cone. Then, the variation of
$Det~D(\alpha )_B$ with respect to $\alpha $ is given by (see, for
example, \cite{APS,Forman}) \begin{equation} \label{DD}\frac d{d\alpha
}\ln \ Det~D(\alpha )_B=\frac d{dz}\left[ \ z\ Tr\{\frac d{d\alpha }\left(
D(\alpha )_B\right) \ D(\alpha )_B^{z-1}\}\right] _{z=0} .  \end{equation}
Note that, under the assumptions made, $\frac d{d\alpha }\left( D(\alpha
)_B\right) $ is a multiplication operator. 
 
Although $J_z(x,t;x,t;\alpha),$ the kernel of $D(\alpha)_B^z$ evaluated at
the diagonal, can be extended to the whole $z$-complex plane as a
meromorphic function, the r.h.s. in (\ref{DD}) cannot be simply written as
the integral over $M$ of the finite part of \begin{equation} tr\{\frac
d{d\alpha }\left( D(\alpha )_B\right) \ J_{z-1}(x,t;x,t;\alpha)\}
\end{equation} at $z=0$ (where $tr$ means matrix trace). In fact,
$J_{z-1}(x,t;x,t;\alpha)$ is in general non integrable in the variable $t$
near $  \partial M$ for $z\approx 0$. 
 
Nevertheless, an integral expression for $\frac d{d\alpha }\ln \
Det~D(\alpha )_B$ will be constructed in the next section from the
integral expression for $Tr(D(\alpha)_B^{z-1})$, holding in a neighborhood
of $z=0$, obtained in the following way \cite{Seeley2}: 
 
if $T>0$ is small enough, the function $j_z(x;\alpha)$ defined as
 \begin{equation} \label{T}j_z(x;\alpha)\ =\int_0^TJ_z(x,t;x,t;\alpha)\ dt
\end{equation} for $Re\ z<1-\nu $, admits a meromorphic extension to {\bf
C} as a function of $z$. So, if $V$ is a neighborhood of $ \partial M$
defined by $t<\epsilon $, with $\epsilon $ small enough,
$Tr(D(\alpha)_B^{z-1})$ can be written as the finite part of
\begin{equation} \label{EI} \int_{M/ V}tr\ J_{z-1}(x,t;x,t;\alpha)\
dxdt+\int_{\partial M}tr\ j_{z-1}(x;\alpha)\ dx\ , \end{equation} where a
suitable partition of the unity is understood. 
 
 \eject
 
\section{Green functions and determinants} 
 
In this section, we will establish an expression for $\frac d{d\alpha }\ln
\ Det[D(\alpha )_B]$ in terms of $G_B(x,t;y,s;\alpha )$  , the Green
function of $D(\alpha )_B$ (i.e., the kernel of the operator $[D(\alpha
)_B]^{-1}).$
 
With the notation of the previous Section, (\ref{DD}) can be rewritten as: 
\begin{equation} \label{DDD}\frac d{d\alpha }\ln \ Det~D(\alpha
)_B=\begin{array}{c} \\ F.P. \\ ^{_{z=0}} \end{array}
 \int_Mtr\left[ \frac d{d\alpha }\left( D(\alpha )_B\right) \
J_{-z-1}(x,t; x,t;\alpha )\right] \ d\bar x\ , \end{equation} where the
r.h.s. must be understood as the finite part of the meromorphic extension
of the integral at $z=0$. 

The finite part of $J_{-z-1}(x,t;x,t;\alpha )$ at $z=0$ does not coincide
with the regular part of $G_B(x,t;y,s;\alpha )$ at the diagonal, since the
former is defined through an analytic extension. 
 
However, we will show that there exists a relation between them, involving
a finite number of Seeley's coefficients. In fact, for boundaryless
manifolds this problem has been studied in \cite{Jour.1}, by comparing the
iterated limits $F.P.\lim \limits_{z\rightarrow -1}\{\lim \limits_{\bar
y\rightarrow \bar x}J_z(x,t;y,s;\alpha)\}$ and $R.P.\lim \limits_{\bar
y\rightarrow \bar x}\{\lim \limits_{z\rightarrow
-1}J_z(x,t;y,s;\alpha)\}=$ $R.P.\lim \limits_{\bar y\rightarrow \bar
x}G_{B}(x,t;y,s;\alpha).$
 
In the case of manifolds with boundary, the situation is more involved
owing to the fact that the finite part of the extension of
$J_z(x,t;x,t;\alpha)$ at $z=-1$ is not integrable near $\partial M$ $.$ (A
first approach to this problem appears in \cite{Jour.2}). Nevertheless, as
mentioned in Section 2, a meromorphic extension of $  \int_0^T
J_z(x,t;x,t;\alpha)dt,$ with $T$ small enough can be performed and its
finite part at $z=-1$ turns to be integrable in the tangential variables.
A similar result holds, {\it afortiori}, for $\int_0^T t^n
J_z(x,t;x,t;\alpha)dt,$ with $n=1,2,3...$
 Then, near the boundary, the Taylor expansion of the function $A_\alpha
=\frac d{d\alpha }D(\alpha )_B$ will naturally appear, and the limits to
be compared are $F.P.\lim \limits_{z\rightarrow -1}\{\lim \limits_{\bar
y\rightarrow \bar x}\int_0^Tt^nJ_z(x,t;y,s;\alpha)dt\} $ and $ R.P.\lim
\limits_{\bar y\rightarrow \bar x}\{\lim \limits_{z\rightarrow
-1}\int_0^Tt^nJ_z(x,t;y,s;\alpha)dt\}=R.P.\lim \limits_{\bar y\rightarrow
\bar x}\int_0^Tt^nG_{B}(x,t;y,s;\alpha)dt.$

\bigskip
 
The starting point for this comparison will be to carry out asymptotic
expansions and to analyze the terms for which the iterated limits do not
coincide (or do not even exist). 
 
An expansion of $G_B(x,t,y,s)$ in $M\backslash \partial M$ in homogeneous
and logarithmic functions of $(\bar x-\bar y)$ can be obtained from (\ref
{AE}) for $\lambda =0$ (see \cite{LibroAmarillo}):  \begin{equation}
\label{?} \begin{array}{c} G_B(x,t,y,s)= \sum_{j=1-\nu
}^0h_j(x,t,x-y,t-s)+M(x,t)\log \vert (x,t)-(y,s)\vert +R(x,t,y,s),
\end{array} \end{equation} with $h_j$ the Fourier transform{\cal \ }${\cal
F}^{-1}(c_{-\nu -j})$ for $j>0$ and $h_0={\cal \ }{\cal F}^{-1}(c_{-\nu
})-\ M(x,t)\log \vert (x,t)-(y,s)\vert .$ The function $M(x,t)$ will be
explicitly computed below (see (\ref{MM})). Our convention for the Fourier
transform is \begin{equation} \label{Fourier} \begin{array}{c}
\displaystyle {\cal F}(f)(\bar \xi )=\hat f(\bar \xi )=\int f(\bar x)\
e^{-i\bar x.\bar \xi }\ d\bar x, \\ \\ \displaystyle\ {\cal F}^{-1}(\hat
f)(\bar x)=f(\bar x)=\dfrac 1{(2\pi )^\nu }\int \hat f(\bar \xi )\
e^{i\bar x.\bar \xi }\ d\bar \xi .  \end{array} \end{equation}
 
For $t>0$, $R(x,t,y,s)$ is continuous even at the diagonal ($y,s)=(x,t)$. 
Nevertheless, \linebreak $R(x,t,y,s)\vert _{(y,s)=(x,t)}$ is not
integrable because of its singularities at $t=0$. On the other hand, the
functions $t^nR(x,t,y,t)$ are integrable with respect to the variable $t$
for $y\neq x$ and $  n=0,1,2,....$An expansion of $\int_0^\infty
t^nR(x,t,y,t)dt$ in homogeneous and logarithmic functions of $(x-y)$ can
also be obtained from ($\ref{AE})$:  \begin{equation} \label{Asn}
\int_0^\infty t^nR(x,t,y,t) dt=\sum_{j=n+2-\nu
}^0g_{j,n}(x,x-y)+M_n(x)\log (\vert x-y\vert )+R_n(x,y) \end{equation}
where $R_n(x,y)$ is continuous even at $y=x$, and $g_{j,n}$ is the Fourier
transform of the (homogeneous extension of) $\int_0^\infty t^n\tilde
d_{-1-j}(x,t,\xi ,t,0)\ dt$, with \begin{equation} \label{`d}\tilde
d_{-1-j}(x,t,\xi ,s,\lambda )=-\int_{\Gamma ^{-}}e^{-is\tau }\
d_{-1-j}(x,t,\xi ,\tau ,\lambda )\ d\tau \end{equation} for $\Gamma ^{-}$
a closed path enclosing the poles of $d_{-1-j}(x,t,\xi ,\tau ,\lambda )$
lying in $\{Im\ \tau >0\}$. 
 
Since $\tilde d_{-1-j}$ is homogeneous of degree $-j$ in (1/t, $\xi
,1/s,\lambda )$, $g_{j,n}$ turns out to be homogeneous of degree $j$ in
$x-y$. 
 
The following technical lemma will be used for the proof of our main
result (Theorem 1): 
 
\bigskip\  
 
{\bf Lemma 1:} {\it Let $a(\xi )$ a function defined on ${\bf R}^\nu $,
homogeneous of degree -$\nu $ for $\vert \xi \vert \geq 1$ and $a(\xi )=0$
for $\vert \xi \vert <1.$ Then its Fourier transform can be written as
\begin{equation} \label{4.1}{\cal \ \ }{\cal F}^{-1}(a(\xi ))(z)=h(z)+\ M\
\frac{\Omega _\nu }{(2\pi )^\nu }\ (\log \vert z\vert ^{-1}+{\cal K}_\nu
)+R(z), \end{equation} where
 
a) $h(z)$ is a homogeneous function of degree $0$, such that $\int_{\vert
z\vert =1}h(z)\ d\sigma _z=0.$ It is given by \begin{equation}
\label{4.3}h(z)={\cal \ \ }{\cal F}^{-1}(P.V.[a(\xi /\vert \xi \vert )-M\
]\vert \xi \vert ^{-\nu })(z).  \end{equation}

b) \begin{equation} \label{4.2}M=\frac 1{\Omega _\nu }\int_{\vert \xi
\vert =1}a(\xi )\ d\sigma _\xi , \end{equation} \\ where $\Omega _\nu
=Area(S^{\nu -1}),$ and ${\cal K}_\nu =\ln 2-\frac 12\gamma +\frac
12\frac{\Gamma ^{\prime }(\nu /2)}{\Gamma (\nu /2)}$ with $ \gamma $ the
Euler's constant. 

\bigskip 
 
c) $R(z)$ is a function regular at $z=0$ with $R(0)=0.$ }

\bigskip

{\bf Proof:} Writing $a(\xi ) = \tilde a(\xi ) + M \vert \xi \vert ^{-\nu
} \chi(\xi)$, with $\tilde a(\xi )$ having zero mean on $\vert \xi \vert =
1$, and $\chi$ the characteristic function of $\vert \xi \vert \geq 1$,
this lemma follows from direct computations , by using the techniques for
Fourier transforms of homogeneous functions (see for instance
\cite{LibroAmarillo}.)

\bigskip
 
 Now, we introduce the main result of this paper. 
 
\bigskip 

{\bf Theorem 1:} {\it Let $M$ be a bounded closed domain in ${\bf R}^\nu $
with smooth boundary $\partial M$ and $E$ a $k$-dimensional complex vector
bundle over $M$. 
 
Let $(D_\alpha )_{B}$ be a family of elliptic differential operators of
first order, acting on the sections of $E$, with a fixed local boundary
condition $B$ on $ \partial M$, and denote by $J_z(x,t;x,t;\alpha)$ the
meromorphic extension of the evaluatio n at the diagonal of the kernel of
$((D_\alpha )_B)^z$. 
 
Let us assume that, for each $\alpha $, $(D_\alpha )_B$ is invertible, the
family is differentiable with respect to $  \alpha ,$ and $\dfrac \partial
{\partial \alpha }(D_\alpha )_{B}f=A_\alpha f$ , with $A_\alpha $ a
differentiable function. 
 
If $V$ is a neighborhood of $\partial M$ defined by $t < \epsilon$ and
$T>0$ small enough, then: 
 
a) } \begin{equation} \label{Cuculiu} \begin{array}{c} \displaystyle
\dfrac \partial {\partial \alpha }\ln \ Det(D_\alpha )_B =
\begin{array}{c} \\ F.P. \\ ^{_{z=-1}} \end{array} \displaystyle \left[
\int_{ \partial M}\int_0^Ttr\left\{ A_\alpha (x,t) \ J_z(x,t;x,t;\alpha)\
\right\} dtdx\right] \\ \\ +\begin{array}{c} \\ F.P. \\ ^{_{z=-1}}
\end{array} \displaystyle \left[ \int_{ M/V}tr\left\{ A_\alpha (\bar x)\
J_{z}(\bar x;\bar x;\alpha)\ \right\}d\bar x\right], \end{array}
\end{equation} {\it where a suitable partition of the unity is understood.
(This expression must be understood as the finite part at $z=-1$ of the
meromorphic extension). 
 
b) For every $\alpha$, the integral $\int_0^TA_\alpha (x,t)\ J_z
(x,t;x,t;\alpha)dt\ $ is a meromorphic function of $z$, for each $x\in
\partial M$, with a simple pole at $z=-1$. Its finite part (dropping, from
now on, the index $\alpha$ for the sake of simplicity) is given by}
\begin{equation} \label{reputamadre} \begin{array}{c} \begin{array}{c} \\
\displaystyle F.P. \\ ^{_{z=-1}} \end{array} \displaystyle \int_0^TA(x,t)\
J_z(x,t;x,t)dt \displaystyle = -\int_0^TA(x,t)\int_{\vert (\xi ,\tau
)\vert =1}\frac i{2\pi }\int_\Gamma \dfrac{\ln \lambda }\lambda \ c_{-\nu
}(x,t;\xi ,\tau ;\lambda )\ d\lambda \ \dfrac{d\sigma _{\xi ,\tau
}}{{(2\pi )^\nu }}\ dt\ \\ \\ \displaystyle +\sum\limits_{l=0}^{\nu -2}
\dfrac{\partial _t^lA(x,0)}{l!}\int_{\vert \xi \vert =1}\int_0^\infty t^l\
\frac i{2\pi }\int_\Gamma \dfrac{\ln \lambda }\lambda \ \tilde d_{-(\nu
-1)+l}(x,t;\xi ,t;\lambda)\ d\lambda \ \ dt\ \dfrac{d\sigma _\xi }{{{(2\pi
)^{\nu -1}}}}\ \\ \\ \displaystyle +\lim \limits_{y\rightarrow x}\left\{
\int_0^TA(x,t)\left[ G_B(x,t;y,t)-\sum\limits_{l=1-\nu
}^0h_l(x,t;x-y,0)\right. \right.  \displaystyle \left. -M(x,t)\
\dfrac{\Omega _\nu }{{(2\pi )^\nu }}\left( \ln \vert x-y\vert ^{-1}+{\cal
K} _\nu \right) \right] dt \\ \\ \displaystyle +\sum\limits_{j=0}^{^{\nu
-2}}\sum\limits_{l=0}^{\nu -2-j} \dfrac{\partial _t^lA(x,0)}{l!}\
g_{j,l-(\nu -2-j)}(x,x-y) \displaystyle +\left. \sum_{l=0}^{^{\nu
-2}}\dfrac{\partial _t^lA(x,0)}{l!}\ M_{\nu -2-l}(x)\ \dfrac{\Omega _{\nu
-1}}{{{(2\pi )^{\nu -1}}}}\left( \ln \vert x-y\vert ^{-1}+{\cal K}_{\nu
-1}\right) \right\} , \end{array} \end{equation} {\it with }
\begin{equation} \label{MM} \begin{array}{c} \displaystyle M(x,t)=\dfrac
1{\Omega _\nu }\int_{\vert (\xi ,\tau )\vert =1}c_{-\nu }(x,t;\xi ,\tau
;0)\ \ d\sigma _{\xi ,\tau } \\ \\ \displaystyle M_j(x)=\dfrac 1{\Omega
_{\nu -1}}\int_{\vert \xi \vert =1}\int_0^\infty t^{\nu -2-j}\ \ \tilde
d_{-1-j}(x,t;\xi ,t;0)\ dt\ d\sigma _\xi , \end{array} \end{equation} {\it
where $\Omega _n=Area(S^{n-1})$, and where $h_{l}$ and $g_{l,n}$ are
related to the Green function $G_B$ as in (\ref{?}) and (\ref{Asn}) }
\begin{equation} \label{hg} \begin{array}{c} \displaystyle h_{1-\nu
+j}(x,t;w,u)
 = {\cal F}_{(\xi ,\tau )}^{-1}\ \left[ c_{-1-j}(x,t;(\xi ,\tau )/\vert
(\xi ,\tau )\vert ;0)\ \vert (\xi ,\tau )\vert ^{-1-j}\right] (w,u), \\ \\
 
\displaystyle h_0(x,t;w,u) = {\cal F}_{(\xi ,\tau )}^{-1}\left[
P.V.\left\{ \left( c_{-\nu }(x,t;(\xi ,\tau )/\vert (\xi ,\tau )\vert
;0)-M(x,t)\right) \ \vert (\xi ,\tau )\vert ^{-\nu }\right\} \right]
(w,u), \\ \\
 
\displaystyle g_{j,l}(x,w)
 = {\cal F}_\xi ^{-1}\left[ \int_0^\infty t^n\ \tilde d_{-1-j}(x,t;\xi
/\vert \xi \vert ,t;0)\ dt\vert \xi \vert ^{-1-j-n}\right] (w), \\ \\
\end{array} \end{equation} \\ {\it with $l = j + n - \nu + 2 $ and }

\begin{equation} \displaystyle g_{j,0}(x,w) ={\cal F}_\xi ^{-1}\left[
P.V.\left[ \int_0^\infty t^{\nu -j-2}\ \tilde d_{-1-j}(x,t;\xi /\vert \xi
\vert ,t;0)\ dt-M_j(x)\right] \vert \xi \vert ^{-(\nu -1)}\right] (w). 
\end{equation} \\ {\it c) The integral $\int_{M\backslash V}tr\left[ A
(\bar x)\ J_{z}(\bar x;\bar x)\right] \ d\bar x $ in the second term in
the r.h.s. of (\ref{Cuculiu}) , is a meromorphic function of $z$ with a
simple pole at $z=-1$. Its finite part is given by } \begin{equation}
\begin{array}{c} \begin{array}{c} \\ F.P. \\ ^{_{z=-1}} \end{array}
\displaystyle \int_{M\backslash V}tr\left[ A(\bar x)\ J_{z}(\bar x;\bar
x)\right] \ d\bar x \displaystyle =\int_{M\backslash V}A (\bar
x)\int_{\vert \bar \xi \vert =1}\dfrac i{2\pi }\int \dfrac{\ln \lambda
}\lambda \ c_{-\nu}(\bar x,\bar \xi ;\lambda )\ d\lambda \dfrac{d\bar \xi
}{(2\pi )^\nu } \\ \\ \displaystyle +\int_{M\backslash V}\lim
\limits_{\bar y\rightarrow \bar x}\ A (\bar x)[G_B(\bar x,\bar
y)-\sum\limits_{l=1-\nu }^0h_l(\bar x,\bar x-\bar y) \displaystyle -M(\bar
x)\dfrac{\Omega _\nu }{(2\pi )^\nu }(\ln \vert \bar x-\bar y\vert
^{-1}+{\cal K}_\nu )]\ d\bar x.  \end{array} \end{equation}

{\bf Proof:} Statement a) is a direct consequence of (\ref{DD}), (\ref{T})
and (\ref{EI}).

 \bigskip
 
In what follows, we will proof the assertion in (\ref{reputamadre}). 
 
We will use, as an approximation to $(D_B-\lambda )^{-1}$, the parametrix
$P_K(\lambda)$ in (\ref{100}). 
 
Thus, we can approximate the kernel $J_z$ of $D_B^z$ by means of the
kernel $L_z^K$ of $\frac i{2\pi }\int_\Gamma \lambda ^z\ P_K(\lambda )\
d\lambda $.  We have \begin{equation} \begin{array}{c} \displaystyle
L_z^K(x,t;y,s)= \sum\limits_\varphi \ \psi (x,t)\left[ \
\sum\limits_{j=0}^K\int_{\bf{R}^\nu }C_{-1-j}(x,t;\xi ,\tau ;z)\
e^{i[(x-y)\xi +(t-s)\tau ]}\ {\dfrac{{d\xi }}{{(2\pi )^{\nu -1}}}}\
{\dfrac{{d\tau }}{{2\pi }}}\right.  \\ \displaystyle \left.
-\sum\limits_{j=0}^K\int_{{\bf R}^{\nu -1}}D_{-1-j}(x,t;\xi ,\tau ;z)\
e^{i(x-y)\xi \ }{\dfrac{{d\xi }}{{(2\pi )^{\nu -1}}}}\right] \ \varphi
(y,s) \end{array} \end{equation}
 with \begin{equation} C_{-1-j}(x,t;\xi ,\tau ;z)=\frac i{2\pi
}\int_\Gamma \lambda ^z\ \theta _2(\xi ,\tau ;\lambda )\ \
c_{-1-j}(x,t;\xi ,\tau ;\lambda )\ d\lambda , \end{equation} and

\begin{equation} \label{DDDD}D_{-1-j}(x,t;\xi ,t;z)\ \equiv \frac i{2\pi
}\int_\Gamma \lambda ^z\theta _1(\xi ,\lambda )\ \tilde d_{-1-j}(x,t;\xi
,t;\lambda )\ d\lambda.  \end{equation} These expressions are, in fact,
analytic functions of $z$ for all complex $z,$ since the singularities of
$\ c_{-1-j}(\lambda )\ $and $  \tilde d_{-1-j}(\lambda )\ $ are in a
compact set in the $\lambda $ plane, for $(x,t;\xi ,\tau )$ in a compact
set. 
 
Since $(D_B-\lambda )^{-1}-P_K(\lambda )$ has a continuous kernel of $O(\
\vert \lambda \vert ^{\nu -K-1})$ for $\lambda \in \Lambda $
\cite{Seeley1}, it turns out that\\ \begin{equation}
R(x,t;y,s;z)=J_z(x,t;y,s)-L_z^K(x,t;y,s) \end{equation} is a continuous
function of $x,t,y,s$ and $z$, and analytic in $z$ for $  Re(z)<0$, if
$K\geq \nu $. Analyzing the last terms in $L_z^K$, we obtain that it is
also true for $K=\nu -1.$ From now on, we call $L_z=L_z^{\nu -1}.$ Then
\begin{equation} \lim _{z\rightarrow -1}\left[ \lim _{(y,s)\rightarrow
(x,t)}(J_z-L_z)\right] =\lim _{(y,s)\rightarrow (x,t)}\left[ \lim
_{z\rightarrow -1}(J_z-L_z)\right] .  \end{equation}\\ Since
\begin{equation} J_{-1}(x,t;y,s)=G_B(x,t;y,s),\ \ \hbox {for}\ (x,t)\neq
(y,s), \end{equation} we have \begin{equation} \label{jl}\lim
_{z\rightarrow -1}(J_z(x,t;x,t)-L_z(x,t;x,t))=\lim _{(y,s)\rightarrow
(x,t)}(G_B(x,t;y,s)-L_{-1}(x,t;y,s)). \\ \end{equation}

\bigskip One can cancel some terms in the equality (\ref{jl}) by studying
the singularities of $L_z(x,t;x,t)$ at $z=-1$ and those of
$L_{-1}(x,t;y,s)$ at $(x,t)=(y,s)$. More precisely: 

\bigskip

{\bf Lemma 2} :

 {\it The following statement holds}
 
\begin{equation} \label{lemma3} \begin{array}{c} \displaystyle \lim
\limits_{z\rightarrow -1}\left[ J_z(x,t;x,t)+\frac 1{(z+1)}\int_{\vert
(\xi ,\tau )\vert =1}c_{-\nu }(x,t;\xi ,\tau ;0)\ \dfrac{d\sigma _{\xi
,\tau }}{(2\pi )^\nu }\right.  \\ \\ \displaystyle +\int_{\vert (\xi ,\tau
)\vert =1}\frac i{2\pi }\int_\Gamma \frac{\ln \lambda }\lambda c_{-\nu
}(x,t;\xi ,\tau ;\lambda )\ d\lambda \ \dfrac{d\sigma _{\xi ,\tau }}{(2\pi
)^\nu } \displaystyle \left. +\sum\limits_{j=0}^{\nu -1}\int_{ {\bf
R}^{\nu -1}}D_{-1-j}(x,t;\xi ,t;z)\ \ {\dfrac{{d\xi }}{{(2\pi )^{\nu -1}
}}}\right] \\ \\ \displaystyle =\lim \limits_{y\rightarrow x}\left[
G_B(x,t;y,t)-\sum\limits_{l=1-\nu }^0h_l(x,t;x-y,0)\right.  \\ \\
\displaystyle -\ M(x,t)\ \dfrac{\Omega _\nu }{(2\pi )^\nu }(\ln \vert
x-y\vert ^{-1}+{\cal K}_\nu ) \displaystyle \left. +\sum\limits_{j=0}^{\nu
-1}\int_{{\bf R}^{\nu -1}}D_{-1-j}(x,t;\xi ,t;-1)e^{i(x-y)\xi }\ \
{\dfrac{{d\xi }}{{(2\pi )^{\nu -1}}}}\right] . \end{array} \end{equation}

\bigskip

{\bf Proof:} Eq.(\ref{lemma3} ) is obtained from (\ref{jl}) in the
following way: 

In the l.h.s., the Fourier transform of the $ C_{-1-j}(x,t;\xi ,\tau
;z)'s$ involved in $L_z(x,t;x,t)$ is written as $\int_{\vert (\xi ,\tau
)\vert \leq1}+\int_{\vert (\xi ,\tau )\vert >1} $, taking into account
their homogeneity for ${\vert (\xi , \tau )\vert >1}$ and analyticity in
z. 

In the r.h.s., the Fourier transforms of
 $C_{-1-j}(x,t;(\xi ,\tau )/\vert (\xi ,\tau )\vert ;-1)\ \ \vert (\xi
,\tau )\vert ^{-1-j}$ is added and subtracted (this gives rise to the
functions $ h_l$ for $l>0$.) Next, compute the Fourier transform of
$C_\nu$ by using Lemma 1. 

Finally, repeated terms are canceled (Notice that, for $\vert (\xi, \tau
)\vert \geq 1$, $ C_{-\nu}(x,t;\xi ,\tau ;z=-1) $ $=$ \linebreak
$c_{-\nu}(x,t;\xi ,\tau ;\lambda=0)$.)

\bigskip
 
The meromorphic extension of the terms involving the coefficients $
C_{-1-j}(x,t;\xi ,\tau ;z)$ in $L_z(x,t;x,t)$ is a consequence of the
previous arguments.  Although $\sum\limits_{j=0}^{\nu -1}\int_{{\bf
R}^{\nu -1}}D_{-1-j}(x,t;\xi ,t;z)\ \ {\frac{{d\xi }}{{(2\pi )^{\nu
-1}}}}$ does not admit, in general, a meromorphic extension, such
extension can be performed for \begin{equation} \int_0^Tt^n\int_{{\bf
R}^{\nu -1}}D_{-1-j}(x,t;\xi ,\tau ;z)\ \ {\frac{{d\xi }}{{(2\pi )^{\nu
-1}}}}\ dt, \end{equation} for $n=0,1,...$ and $j=0,1,2,...$ (see
\cite{Seeley2} and the next lemma.) Then, in order to prove b) in the
Theorem, we first establish a technical result obtained from the
fundamental estimate \begin{equation} \label{4.11}\vert t^n\partial _\xi
^\alpha \tilde d_{-1-j}(x,t,\xi ,s;\lambda )\vert \leq Ce^{-c(t+s)(\vert
\xi \vert +\vert \lambda \vert )}(\vert \xi \vert +\vert \lambda \vert
)^{-j-n-\vert \alpha \vert }, \end{equation} for $t,s>0$ , $\lambda \in
\Lambda $, due to R.T. Seeley \cite{Seeley1}: 

\bigskip

{\bf Lemma 3:} 
 
{\it For $D_{-1-j}(x,t;\xi ,t;z)$ as in (\ref {DDDD}) it holds: 
 
i) If $r(x,t)$ is a function satisfying $\vert r(x,t)\vert \leq Ct^{n}$
for $0<t<T$, $n\in {\bf N,}$ $T>0,$ \begin{equation}
\label{4.12}\int_0^Tr(x,t)\int_{{\bf R}^{\nu -1}}D_{-1-j}(x,t;\xi ,t;z)\
e^{i(x-y)\xi }\ {\frac{{d\xi}}{{(2\pi)^{\nu -1}}}} \ dt \end{equation} is
an absolutely convergent integral for $Re(z)<j+n-\nu +1.$ As a
consequence, it is an analytic function of $z$ in this region, and it is
continuous in all the variables ($x,y,z).$
 
ii) If $x\neq y$, (\ref{4.12}) is an absolutely convergent integral for
all $z\in {\bf C,\ }$and so no analytic extension is needed out of the
diagonal. 
 
iii) \begin{equation} \label{4.13}\int_0^\infty t^nD_{-1-j}(x,t;\xi ,t;z)\
\ dt \end{equation} is an homogeneous function of $\xi $ for $\vert \xi
\vert \geq 1$, of degree $  z-j-n$, analytic in $z$ for $Re(z)<j+n$ and
then \begin{equation} \begin{array}{c} \label{4.14}\displaystyle
\int_{{\bf R}^{\nu -1}}\int_0^\infty t^nD_{-1-j}(x,t;\xi ,t;z)\ \ dt\
{\frac{{d\xi }}{{(2\pi )^{\nu -1}}}}

\displaystyle =\alpha _j^n(x;z) +\frac 1{z-j-n+\nu -1}\beta _j^n(x;z)
\end{array} \end{equation} with \begin{equation} \label{4.15}
\begin{array}{c} \displaystyle \alpha _j^n(x;z)=\int_{\vert \xi \vert \leq
1}\int_0^\infty t^nD_{-1-j}(x,t;\xi ,t;z)\ \ dt\ {\dfrac{{d\xi }}{{(2\pi
)^{\nu -1}}}} \\ \\ \displaystyle \beta _j^n(x;z)=\int_{\vert \xi \vert
=1}\int_0^\infty t^nD_{-1-j}(x,t;\xi ,t;z)\ \ dt\ {\dfrac{{d\xi }}{{(2\pi
)^{\nu -1}}}} \end{array} \end{equation} analytic functions of $z$ for
$Re(z)<j+n$. 
 
iv) \begin{equation} \label{4.16}\int_{{\bf R}^{\nu -1}}\int_T^\infty
t^nD_{-1-j}(x,t;\xi ,t;z)\ e^{i(x-y)\xi }\ dt\ {\frac{{d\xi}}{{(2\pi)^{\nu
-1}}}} \end{equation} is an entire function of $z$, continuous in
$(x,y,z).$ }

{\bf Proof:} It follows from the homogeneity and analyticity properties of
the functions $\tilde d_{-1-j}$ and the estimate (\ref{4.11}). 
 
\bigskip
 
Now, to get part b) of the theorem, we study the limits $\lim
\limits_{z\rightarrow -1}\int_0^TA(x,t)\ R(x,t;z)\ dt$ and \linebreak
$\lim \limits_{y\rightarrow x}\int_0^TA(x,t)\ S(x,t;y,t)\ dt$, where
$R(x,t;z)$ and $S(x,t;y,t)$ denote the expressions appearing in the limits
on the l.h.s. and r.h.s. of (\ref{lemma3}) respectively.  By considering
the expansion of $A(x,t)$ in powers of $t$, we obtain: 

\bigskip
 
{\bf Lemma 4: }{\it If $A(x,t)$ has $\nu -1-j$ continuous derivatives in
the variable $t,$ $t\geq 0,$ then
 
i) For $\nu -1-j>0$,
 \begin{equation} \begin{array}{c} \displaystyle \int_0^TA(x,t)\int
D_{-1-j}(x,t;\xi ,t;z)\ {\dfrac{{d\xi }}{{(2\pi )^{\nu -1}}}}\ dt=\psi
_j(x,z)\ \\ \\ \displaystyle -\dfrac 1{z+1} \dfrac{\partial _t^{\nu
-j-2}A(x,0)}{(\nu -j-2)!}\ \int_{\vert \xi \vert =1}\int_0^\infty t^{\nu
-j-2}D_{-1-j}(x,t;\xi ,t;-1)\ dt\ \dfrac{d\sigma _\xi }{{{(2\pi )^{\nu
-1}}}} \\ \\ \displaystyle \ -\dfrac{\partial _t^{\nu -j-2}A(x,0)\ }{(\nu
-j-2)!}\int_{\vert \xi \vert =1}\int_0^\infty t^{\nu -j-2}\partial
_zD_{-1-j}(x,t;\xi ,t;-1)\ dt\dfrac{  d\sigma _\xi }{{{(2\pi )^{\nu
-1}}}}, \end{array} \end{equation} with $\psi _j(x,z)$ an analytic
function of $z$ for $Re(z)<0.$
 
Moreover,
 \begin{equation} \begin{array}{c} \displaystyle \int_0^TA(x,t)\int
D_{-1-j}(x,t;\xi ,t;-1)\ e^{i(x-y)\xi }\ {\dfrac{{d\xi }}{{(2\pi )^{\nu
-1}}}}\ dt \displaystyle =\varphi _j(x,y)\ \\ \\
\displaystyle+\sum\limits_{ \QATOP{n=0}{l=j+n-\nu +2}}^{\nu -j-2}\
\dfrac{\partial _t^nA(x,0)}{n!}\ g_{j,l}(x,x-y)\ \displaystyle
+\dfrac{\partial _t^{\nu -j-2}A(x,0)\ }{(\nu -j-2)!}\ \ M_j(x)\ \dfrac{
\Omega _{\nu -1}}{{{(2\pi )^{\nu -1}}}}(\ln \vert x-y\vert ^{-1}+{\cal
K}_{\nu -1}), \end{array} \end{equation} where $\varphi _j(x,y)$ is a
continuous function. 
 
ii) For $\nu -1-j=0$, \begin{equation} \int_0^TA(x,t)\int D_{-1-j}(x,t;\xi
,t;z)\ {\frac{{d\xi}}{{(2\pi)^{\nu -1}}}} \ dt=\psi _j(x,z) \end{equation}
is an analytic function of $z$ for $Re(z)<0$, and \begin{equation}
\int_0^TA(x,t)\int D_{-1-j}(x,t;\xi ,t;-1)\ e^{i(x-y)\xi }\ {\dfrac{{d\xi
}}{{(2\pi )^{\nu -1}}}}\ dt =\varphi _j(x,y) \end{equation} is a
continuous function. 
 
iii) For all $j$ } \begin{equation} \label{510}\lim _{z\rightarrow -1}\
\psi _j(x,z)=\lim _{y\rightarrow x}\ \varphi _j(x,y) \end{equation}

\bigskip {\bf Proof:} For analyzing the expression $\int_0^TA(x,t)\int
D_{-1-j}(x,t;\xi ,t;z)\ d\xi \ dt $ for $z\rightarrow -1$, we develop $A$
in powers of $t$ and apply Lemma 3. 

For the integral $\int_0^TA(x,t)\int D_{-1-j}(x,t;\xi ,t;-1)\ e^{i(x-y)\xi
}\ d\xi \ dt $, we expand $A$, use Lemma 3 and evaluate the Fourier
transforms with the same technique as in Lemma 2, finding the singular
terms given by the functions $g_{j,l}$ and $\ln \vert x-y\vert $. 

\bigskip
 
Finally, in order to get part b) of Theorem 1 we write the equality in
Lemma 2 as \begin{equation} \lim _{z\rightarrow -1}\ R(x,t;z)=\lim
_{y\rightarrow x}\ S(x,y,t) \end{equation} and evaluate the integrals
$\int_0^TA(x,t)\ R(x,t;z)\ dt$ and $  \int_0^TA(x,t)\ S(x,y,t)\ dt.$

For the first one, we have \begin{equation} \label{18} \begin{array}{c}
\displaystyle \int_0^TA(x,t)\left[ J_z(x,t;x,t)+\frac 1{z+1}\int_{\vert
(\xi ,\tau )\vert =1}c_{-\nu }(x,t;\xi ,\tau ;0) \dfrac{\ d\sigma _{\xi
,\tau }}{(2\pi )^\nu }\right.  \\ \\ \displaystyle +\left. \int_{\vert
(\xi ,\tau )\vert =1}\frac i{2\pi }\int \dfrac{\ln \lambda }\lambda
c_{-\nu }(x,t;\xi ,\tau ;\lambda )\ d\lambda \ \dfrac{\ d\sigma _{\xi
,\tau }}{(2\pi )^\nu }\right] \ dt \\ \\ \displaystyle
=-\sum\limits_{j=0}^{\nu -1}\int_0^TA(x,t)\ \int D_{-1-j}(x,t;\xi ,t;z)\
{\dfrac{{d\xi }}{{(2\pi )^{\nu -1}}}}\ dt \displaystyle +\int_0^TA(x,t)\
R(x,t;z)\ dt \\ \\ \displaystyle =\sum\limits_{j=0}^{\nu -2}\frac 1{z+1}
\dfrac{\partial _t^{\nu -j-2}A(x,0)}{(\nu -j-2)!}\int_{\vert \xi \vert
=1}\int_0^\infty t^{\nu -j-2}D_{-1-j}(x,t;\xi ,t;-1)\ \ dt\ \dfrac{\
d\sigma _\xi }{(2\pi )^{\nu -1}} \\ \\ \displaystyle
+\sum\limits_{j=0}^{\nu -2} \dfrac{\partial _t^{\nu -j-2}A(x,0)}{(\nu
-j-2)!}\int_{\vert \xi \vert =1}\int_0^\infty t^{\nu -j-2}\partial
_zD_{-1-j}(x,t;\xi ,t;-1)\ \ dt\ \dfrac{\ d\sigma _\xi }{(2\pi )^{\nu -1}}
\\ \\ \displaystyle -\sum\limits_{j=0}^{\nu -1}\psi
_j(x,z)+\int_0^TA(x,t)\ \ R(x,t;z)\ dt.  \end{array} \end{equation} For
the integral involving $\ S(x,y,t),$ we have \begin{equation} \label{19}
\begin{array}{c} \displaystyle \int_0^TA(x,t)\left[ G_B(x,t;y,t)\right. 
\left. -\sum\limits_{l=-(\nu -1)}^0h_l(x,t;x-y;0)-\ M(x,t) \dfrac{\Omega
_\nu }{(2\pi )^\nu }\left( \ln \vert x-y\vert ^{-1}+{\cal K}_\nu \right) \
\right] dt \\ \\ \displaystyle =-\sum\limits_{j=0}^{\nu -1}\int_0^TA(x,t)\
\int D_{-1-j}(x,t;\xi ,t;-1)\ e^{i(x-y)\xi }\ {\dfrac{{d\xi }}{{(2\pi
)^{\nu -1}}}}\ dt \displaystyle +\int_0^TA(x,t)\ \ S(x,y,t)\ dt \\ \\
\displaystyle =-\sum\limits_{j=0}^{\nu -2}\sum\limits_{n=0}^{\nu -j-2}
\dfrac{\partial _t^nA(x,0)}{n!}g_{j,j+n+\nu -2}(x,x-y) \\ \\ \displaystyle
-\sum\limits_{j=0}^{\nu -2} \dfrac{\partial _t^{\nu -j-2}A(x,0)}{(\nu
-j-2)!}\ \ M_j(x)\dfrac{\Omega _{\nu -1}}{(2\pi )^{\nu -1}}\left( \ln
\vert x-y\vert ^{-1}+{\cal K}_{\nu -1}\right) \displaystyle
-\sum\limits_{j=0}^{\nu -1}\varphi _j(x,y)+\int_0^TA(x,t)\ \ S(x,y,t)\ dt. 
\end{array} \end{equation} Then, taking into account that the last terms
in (\ref{18}) and (\ref{19}) satisfy
 \begin{equation} \begin{array}{c} \displaystyle \lim
\limits_{z\rightarrow -1}\left( -\sum\limits_{j=0}^{\nu -1}\psi
_j(x,z)+\int_0^TA(x,t)\ \ R(x,t;z)\ dt\right) \displaystyle =\lim
\limits_{y\rightarrow x}\left( -\sum\limits_{j=0}^{\nu -1}\varphi
_j(x,y)+\int_0^TA(x,t)\ \ S(x,y,t)\ dt\right) , \end{array} \end{equation}
we obtain part b) of Theorem 1. 
 
The proof of c) is similar to the one of b), and even simpler because in
this case the parametrix in (\ref{100}) does not include terms of the form
Op'($\theta_1 ,\tilde{d}_{-1-j}).$ \qed

\bigskip Eq.(\ref{reputamadre}) looks cumbersome, but it is not so
complicated. In fact, all terms can be systematically evaluated. Moreover,
the terms containing $h_l$ subtract the singular part of the Green
function in the interior of the manifold (see (\ref{?})) and can, thus, be
easily identified from the knowledge of $G_B$. $R(x,t,y,t)$, the regular
part so obtained, is still nonintegrable near the boundary. Those terms
containing $g_{j,l}$ subtract the singular part of the integrals
$\int_0^T\ t^n\ R(x,t,y,t)\ dt$ (see (\ref{Asn})). Finally, the terms
containing $c_{-\nu }$ and $\tilde{d}_{-\nu +1} $ arise as a consequence
of having replaced an analytic regularization by a {\it point splitting }
one. 
 
Even though Seeley's coefficients $c$ and $\tilde d$ are to be obtained
through an iterative procedure, which can make their evaluation a tedious
task, in some cases of physical interest only the few first of them are
needed.  An example of this fact is the calculation we performed in
\cite{porfin}. There, we considered the determinant of the Dirac operator
$D=\ \not \!\!\!i\partial +\not \!\!A$ acting on Dirac fermions defined on
a two dimensional disk, under rather general local elliptic boundary
conditions. In that computation we needed only two Seeley's coefficients.

 \bigskip
 
{\it Acknowledgments}.  We are grateful to R.T. Seeley for useful
comments.  \bigskip

\bigskip
\eject

\end{document}